\documentclass[RNAAS]{aastex631}

\shorttitle{The Minimal Astration Hypothesis}
\shortauthors{Mattsson}
\graphicspath{{./}{figures/}}

\begin{document}

\title{The Minimal Astration Hypothesis -- a Necessity for Solving the Dust Budget Crisis?}

\author[0000-0003-2670-2513]{Lars Mattsson}
\affiliation{Nordita, KTH Royal Institute of Technology and Stockholm University,\\ 
Hannes Alfv\'ens v\"ag 12,\\ 
SE-106 91 Stockholm, Sweden}



\begin{abstract}
Assuming that gas and dust separate in the interstellar medium (ISM) so that high-density regions, where stars can form, are almost devoid of dust, the amount of metals being removed from the ISM can be significantly reduced (minimized astration). Here, it is shown by simple analytical models that this may increase the total metal budget of a galaxy considerably. It is suggested that these extra metals may increase the mass of dust such that the "dust budget crisis", i.e., the fact that there  seems to be more dust at high redshifts than can be accounted for, can be ameliorated. Reducing the amount of astration, the metal budget can be more than doubled, in particular for systems that evolve under continuous gas accretion.
\end{abstract}



\section{Introduction: The Dust Budget Crisis} \label{sec:intro}
Cosmic dust is ubiquitous in the Universe and has a profound effect the light we observe, as well as regulating many important processes from star and planet formation to mass loss in evolved stars. Forming dust grains is not easy and requires that strict combinations of temperature ($\lesssim 2000$\,K), density, and composition are achieved and maintained, while destroying dust is easy via a number of processes, where a combination of shattering and sputtering seems to be most efficient and implies an unexpectedly high rate of destruction \citep[see][and references therein]{Kirchschlager22}. About half of the mass of all metals in the present-day Galaxy are in the form of dust grains, but in  high-redshift galaxies in particular, there seem to be more dust than can be accounted for \citep[see, e.g.,][]{Dwek07,Mattsson11b,Rowlands14b}. This is known as the “dust-budget crisis” (DBC), which is one of the greatest mysteries in contemporary astronomy. 

\section{The Minimal Astration Hypothesis} \label{sec:results}
Regardless of how fast, or by which formation channel, gas-phase metals are converted into dust grains, a strict upper limit to the dust mass is, of course, the {\it total mass of metals}. Metals in interstellar gas are consumed, and removed {\bf from} the matter cycle, when stars form, a.k.a. {\it astration}. The amount of metals astrated is proportional to the metallicity of the star-forming gas. Thus, if metals could be removed from this gas, the overall level of astration would be reduced and the metal budget increased. Here, it is suggested that this can be modeled by a single parameter $\eta$. This ``astration-efficiency parameter'' is a number between zero (no astration) and one (astration in a well-mixed ISM). The underlying hypothesis is based on the fact that dust and gas tend to decouple, in particular when external forces like gravity and radiation pressure are added \citep{Mattsson22,Sandin20,Mattsson19a,Hopkins16}. Larger grains tend to decouple more and they also contain most of the interstellar dust mass ($\sim 90\%$). This means a significant amount of metals may be uncorrelated with the gas and when stars form, these dust grains are not astrated and subsequently driven away by radiation pressure. If such a processes is allowed to {\it minimize the astration}, the DBC can be ameliorated. A related hypothesis was suggested by \citet{Forgan17}, where planet-formation processes were added to the calculations, but the reduction of the amount of dust/metals astrated was deemed insufficient to solve the DBC. For that, a {\it minimal astraion hypothesis} appears to be needed. 

\begin{figure}
  \resizebox{\hsize}{!}{
   \includegraphics{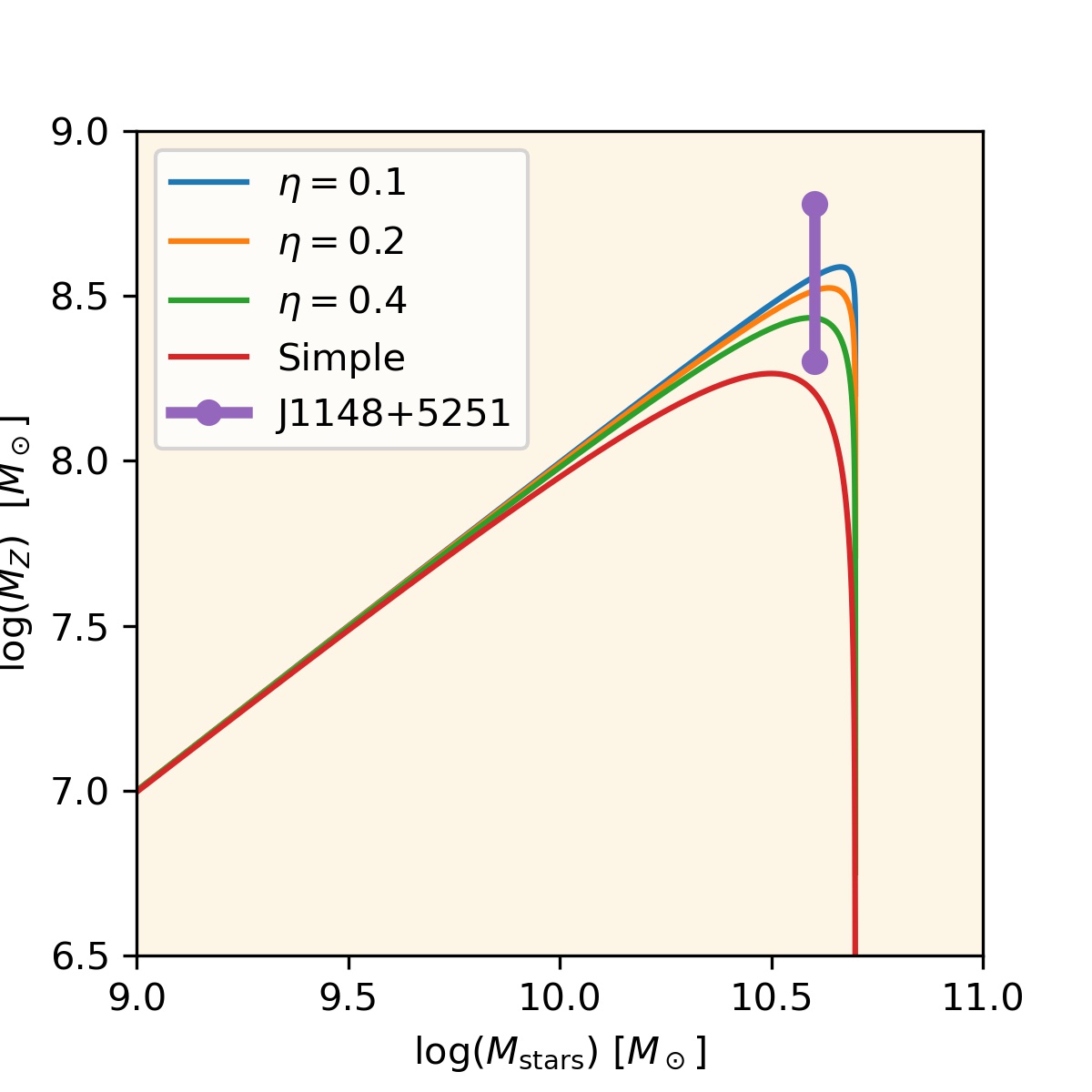}
   \includegraphics{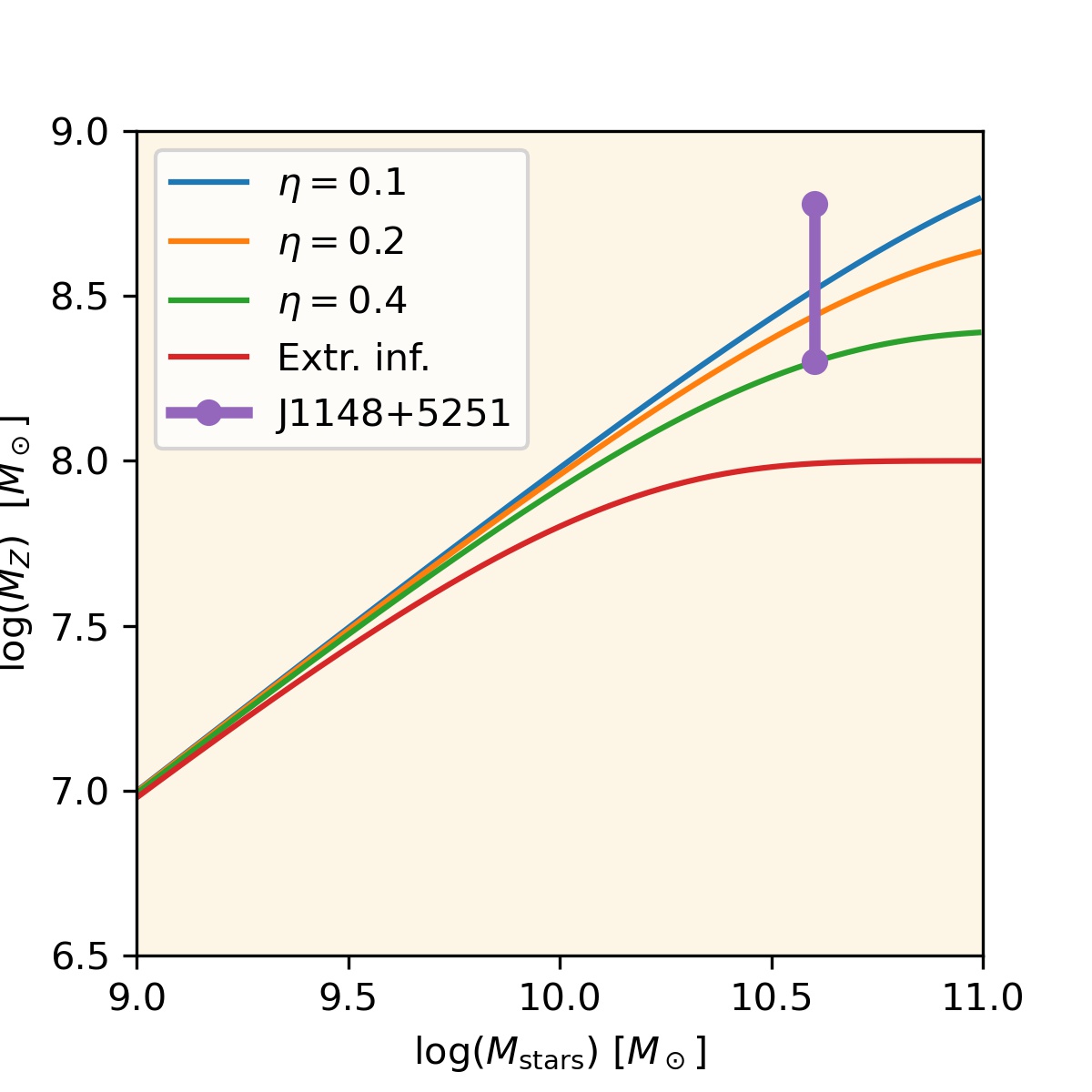}
   }
  \caption{\label{fig:metals} Evolution of the total amount of metals for various astration efficiencies. Left panel shows examples of closed box models, while the right panel shows examples of ``extreme infall'' models. The range of dust-mass estimates for thes quasar-host galaxy J1148+5251 is shown for reference. }
  \end{figure}

A simple closed-box model of chemical evolution with a variable level of astration can be formulated as
\begin{equation}
\label{eq:dmzdt}
    {dM_Z\over dt} = y\,{dM_{\rm s}\over dt} -\eta\,{M_Z\over M_{\rm g}}{dM_{\rm s}\over dt},
\end{equation}
\begin{equation}
\label{eq:dmgdt}
    {dM_{\rm g}\over dt} = -{dM_{\rm s}\over dt},
\end{equation}
where $M_Z$, $M_{\rm g}$, $M_{\rm s}$ are the masses of metals, gas and stars, respectively and $y$ is the effective metal yield. In the following, it is assumed that $y=0.01$, which is a commonly used standard value. Combination of eqns. (\ref{eq:dmzdt}) and (\ref{eq:dmgdt}) yields
\begin{equation}
\label{eq:dzdt}
    M_{\rm g}\,{dZ\over dt} = y\,{dM_{\rm s}\over dt} +(1-\eta)\,Z\,{dM_{\rm s}\over dt},\quad Z = {M_Z\over M_{\rm g}},
\end{equation}
which has the formal solution 
\begin{equation}
\label{eq:cboxeta}
   Z(t) = {y\over \eta-1}\left[1-\left({M_{\rm g}(t)\over M_{\rm tot}} \right)^{\eta-1}\right], \quad 0\le \eta <1.
\end{equation}
\begin{equation}
    Z(t) = y\,\ln\left({M_{\rm tot}\over M_{\rm g}(t)}\right), \quad \eta=1.
\end{equation}
For $0\le \eta <0$, $Z = y/(\eta-1)$ if $M_{\rm g}/M_{\rm tot} \ll 1$, while $Z \to \infty$ in case $\eta = 1$ (no reduction of astration). Example models are shown in the left panel of Fig. \ref{fig:metals}.

A simple model with accretion of pristine gas (``extreme infall model'') can be created by assuming the gas consumption rate and the gas accretion rate cancel and $M_{\rm g}$ remains constant after some initial phase of galaxy formation. Eq. (\ref{eq:dzdt}) would then have an additional term $-Z\,(dM_{\rm s}/dt)$ on the right-hand-side and the solution 
\begin{equation}
\label{eq:einfeta}
   Z(t) = {y\over \eta}\left[1-\exp\left(-\eta\,{M_{\rm s}(t)\over M_{\rm g}} \right)\right], \quad 0 < \eta \le 1.
\end{equation}
\begin{equation}
   Z(t) = y\,{M_{\rm s}(t)\over M_{\rm g}}, \quad \eta = 0.
\end{equation}
For $0 < \eta \le 0$, $Z = y/\eta$ if $M_{\rm s}/M_{\rm g} \gg 1$. Thus, for a given $y$ and $\eta \le 1/2$, the accretion model implies a higher maximal metallicity than the closed box model. Example models are shown in the right panel of Fig. \ref{fig:metals}.

\bigskip
\section{Discussion and Conclusions}
The minimal astration hypothesis rests upon the idea that a very significant fraction of metals is removed from star-forming regions, which requires that a considerable fraction of metals are incorporated in dust in the first place. In the present-day Galaxy, half of the mass of metals is in dust. If most of the dust mass is removed from ISM regions undergoing gravitational collapse, as a result of dynamical decoupling of gas and dust and radiation pressure due to newly formed stars, it is reasonable to claim that $\eta \sim 0.5$. The simple models presented in the previous section imply that the maximal metal budget would double in such a case (see eqns. \ref{eq:cboxeta} and \ref{eq:einfeta}) and with it also the total dust budget. Moreover, emission-line based metallicity measurements will reflect the composition in the star-forming regions and thus imply metallicities below the case of ``normal'' astration ($\eta = 1$), which may explain why some galaxies appear to have even dust-to-metals ratios above unity. A well-known example of a galaxy which is regarded as problematic, is the quasar host J1149+5251. In Fig. \ref{fig:metals} the range of dust masses inferred from observations are shown at a stellar mass $M_{\rm stars}\approx 4.0\cdot 10^{10}\,M_\sun$ resulting from a total baryon mass $M_{\rm tot} = 5\cdot 10^{10}\,M_\sun$ and an estimated gas mass $M_{\rm gas}\sim 10^{10}\,M_\sun$ \citep[see][and references therein]{Mattsson11b,Dwek07}, which are also the masses used as constraints in the example models. The conventional``extreme infall model'' (red line, right panel in Fig. \ref{fig:metals}) is far from producing a sufficient amount of metals, a situation which changes drastically with a lower $\eta$.

Both interstellar dust formation and dust-gas separation may have been enhanced at early epochs, when galaxies are in a formative state and may display more vigorous gas dynamics due to mergers as well as high supernova rates. It is fair to assume that the dust-to-metals ratios were then quite high and the dust-gas separation was generally more pronounced. In such cases, $\eta < 0.5$ is expected, which suggests a metal budget several times larger than that for $\eta = 1$ (see Fig. \ref{fig:metals}). The DBC could then be resolved. However, observed stellar abundances and dust emission in molecular clouds in the present-day Universe are severe constraints, which suggest that $\eta$ is likely closer to one at present.



\bibliography{dust_refs}{}
\bibliographystyle{aasjournal}



\end{document}